\documentclass[journal=jacsat, manuscript=article]{achemso}

\usepackage[version=3]{mhchem} 
\usepackage{amsmath}
\usepackage[T1]{fontenc}
\usepackage[english]{babel}

\author{Grace H. Chen}
\altaffiliation{These authors contributed equally to this work.}
\affiliation[Harvard University]{Applied Physics (SEAS), Harvard University, Cambridge, MA 02138, USA}

\author{Anchita Addhya}
\altaffiliation{These authors contributed equally to this work.}
\affiliation[University of Chicago]{Pritzker School of Molecular Engineering, University of Chicago, Chicago, IL 60637, USA}

\author{Ian N. Hammock}
\altaffiliation{These authors contributed equally to this work.}
\affiliation[University of Chicago]{Pritzker School of Molecular Engineering, University of Chicago, Chicago, IL 60637, USA}

\author{Philip Kim}
\affiliation[Harvard University]{Department of Physics, Harvard University, Cambridge, MA 02138, USA}

\author{Alexander A. High}
\email{ahigh@uchicago.edu}
\affiliation[University of Chicago]{Pritzker School of Molecular Engineering, University of Chicago, Chicago, IL 60637, USA}

\title
  {Extending exciton and trion lifetimes in MoSe$_{2}$ with a nanoscale plasmonic cavity}

\begin{document}

\section{Abstract}
Excitons in transition metal dichalcogenides (TMDs) have extremely short, picosecond-scale lifetimes which hinders exciton thermalization, limits the emergence of collective coherence, and reduces exciton transport in optoelectronic devices. In this work, we explore an all-optical pathway to extend exciton lifetimes by placing MoSe$_2$ in a deep-subwavelength Fabry-Pérot silver cavity. The cavity structure is designed to suppress radiative recombination from in-plane optical dipoles, such as bright excitons and trions. We observe a consistent decrease in photoluminescence (PL) linewidths of excitons and trions ($\approx$ 1 nm), along with a corresponding lifetime increase ($\approx$10 ps). We confirm the experimental observations arise purely from exciton-cavity interactions---etching back the top silver layer returns the PL linewidth and lifetimes return to their original values. Our study offers a pathway to engineer excited state lifetimes in 2D materials which can be utilized for studies of optically dark excitons and have potential applications for novel optoelectronic devices.

\section{Main Text}
Monolayer transition metal dichalcogenides (TMDs) are an exciting class of material for optoelectronic applications due to their direct band-gap optical transitions, strong spin-orbit coupling, and large oscillator strength \cite{xiao_coupled_2012}\cite{mak_control_2012}\cite{mak_photonics_2016}\cite{manzeli_2d_2017}. Exciton (electron-hole pairs) and trion or polaron (charged excitons) emission in TMDs has been demonstrated to be gate-tunable\cite{mak_tightly_2013}\cite{sidler_fermi_2017}. Additionally, TMDs may be stacked in various heterostructures forming long-lived permanent dipole interlayer excitons for studies of Bose-Einstein condensation\cite{wang_evidence_2019}\cite{ma_strongly_2021} \cite{qi_thermodynamic_2023}\cite{qi_perfect_2025}\cite{nguyen_perfect_2025}, light-emitting diodes\cite{jauregui_electrical_2019}\cite{joe_electrically_2021}, and coherent lasing\cite{paik_interlayer_2019}. However, these approaches to extending lifetimes, based on spatial separation of electrons and holes, reduce the exciton binding energy and impact performance at elevated temperatures. In contrast to their long-lived \textit{inter}-layer counterparts, \textit{intra}-layer excitons are short-lived, with radiative lifetimes measured on the scale of several picoseconds\cite{wang_valley_2014}\cite{robert_exciton_2016}. Moreover, the trion, or charged exciton, lifetime has been measured to be $\sim 20$ ps in samples of comparable crystal quality\cite{wang_valley_2014}\cite{robert_exciton_2016}. Increasing the radiative lifetimes of excitons and trions in monolayer transition metal dichalcogenides (TMDs) is crucial for enhancing their potential applications in optoelectronics and valleytronics. Longer lifetimes allow for more efficient manipulation and utilization of these quasiparticles, which are essential for valley-polarized emission and coherent control over optical transitions, while maintaining their exceptionally large binding energy\cite{chernikov_exciton_2014}. Additionally, selective control over the emission of the in-plane bright exciton and out-of-plane dark exciton allows for the detection and study of the higher energy, optically dark exciton in MoSe$_2$ \cite{zhou_probing_2017}\cite{qian_probing_2024}. 

In this work, we explore direct suppression of radiative recombination in a nanoscale cavity as a pathway to generating long-lived intra-layer excitons and trions in monolayer TMDs. Several recent studies have explored the strong coupling between excitons and cavity photons (exciton-polaritons) in moir\'e TMD heterostructures integrated with planar microcavities\cite{zhang_van_2021}\cite{shan_brightening_2022}. In the weak-coupling regime, exciton emission from a cavity may be enhanced or suppressed by the Purcell effect \cite{chen_engineering_2022}\cite{horng_engineering_2019}. In this regime, experiments have shown that when a single mirror is placed at a node ($n\lambda$) or anti-node ($(n+\frac{1}{2})\lambda$), where $\lambda$ is the exciton wavelength and $n$ is an integer, the radiative emission from excitons can be suppressed or enhanced, respectively \cite{horng_engineering_2019} \cite{rogers_coherent_2020}\cite{ren_control_2023}. In particular, there is a thickness transition below which the in-plane electrical field of the cavity mode vanishes, eliminating radiative pathways for bright excitons and trions. Calculations of the radiative behavior of excitons with an \textit{in-plane} optical dipole moment at the center of a tightly-confined ($d<\lambda/2$) planar Fabry-Pérot cavity with metallic interfaces show that emission may be suppressed by several orders of magnitude compared to the spontaneous emission rate without a cavity, corresponding to a significant enhancement of the exciton lifetime\cite{chen_engineering_2022}. In contrast, an out-of-plane optical dipole in the same metallic Fabry-Pérot cavity sees a significant enhancement of its spontaneous emission rate. This distinction between the behavior of in-plane and out-of-plane dipoles in a metallic Fabry-Pérot cavity is not the case for typical 1-dimensional all-dielectric cavities, where the behavior of s- and p-polarized light is identical\cite{joannopoulos_photonic_2011}. 

\begin{figure}[H]
\centerline{\includegraphics{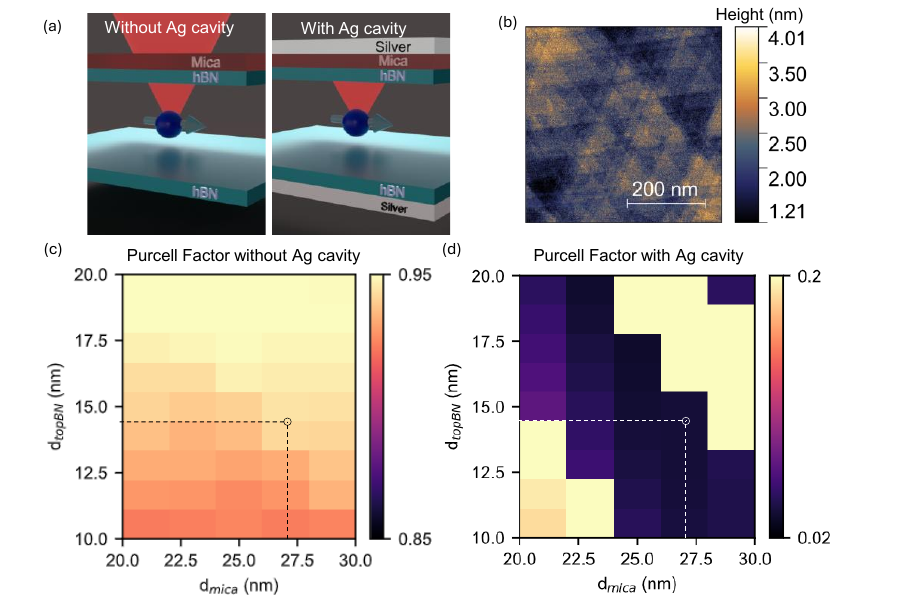}}
\caption{MoSe$_2$ Cavity simulations and fabrication. (a) Schematic of  emission suppression with Ag cavity. (b) Single-crystalline Ag growth on mica substrate. (c,d) Simulated Purcell factors for identifying target thicknesses of top hBN (d\(_{topBN}\)) and mica (d\(_{mica}\)) layer for maximum suppression of PL emission (c) without Ag cavity and (d) with Ag cavity. Simulations used a bottom hBN layer of 48nm.}\label{Figure1}
\end{figure}

To demonstrate the effect of the planar cavity, we use finite-difference time-domain (FDTD) simulations to calculate the Purcell factor of an in-plane point dipole placed in a cavity composed of a bottom layer of silver, hexagonal-boron nitride (hBN) and mica dielectric spacers, and a top layer of silver, illustrated in Figure \ref{Figure1}a. Although it has been shown that excitons in 2-dimensional systems are delocalized and may be better described by a planar dipole description \cite{rogers_coherent_2020} \cite{chen_engineering_2022}, the point dipole model presents an upper-bound for the suppression of radiative emission. Moreover, the planar dipole model coincides with the point dipole model at higher temperatures\cite{chen_engineering_2022}. We find that without the top silver, the emission is negligibly suppressed, while with the addition of the silver cavity, the emission can be suppressed by almost two orders of magnitude (Figure \ref{Figure1}c,d). We found flakes of hBN and mica that corresponded to the thicknesses that yielded the minimum Purcell factor in these simulations. 

We couple excitons in TMDs to the cavity by placing hBN-encapsulated MoSe$_2$ in a planar silver cavity (SI Figure 1). The bottom silver mirror is nominally single-crystalline and grown on a lattice-matched muscovite mica substrate. An atomic force microscopy (AFM) image of the silver surface is shown in Figure \ref{Figure1}b, where the RMS roughness is $<1$nm. Moreover, the AFM image shows that the silver film is continuous with no grain boundaries in the scan area. We identify the triangular pattern of the silver surface to be atomic layer terraces, indicating that the surface is atomically smooth\cite{cheng_epitaxial_2019}. The heterostructure was fabricated via dry transfer method and is composed of a monolayer of MoSe$_2$ encapsulated by hBN with a top layer of exfoliated mica\cite{frisenda_naturally_2020}\cite{castellanosgomez_atomically_2011}. The top layer of exfoliated mica allows us to deposit a smoother film of silver on top of the stack compared to silver deposited directly on hBN (SI Figure 2). 

\begin{figure}[H]
\centerline{\includegraphics{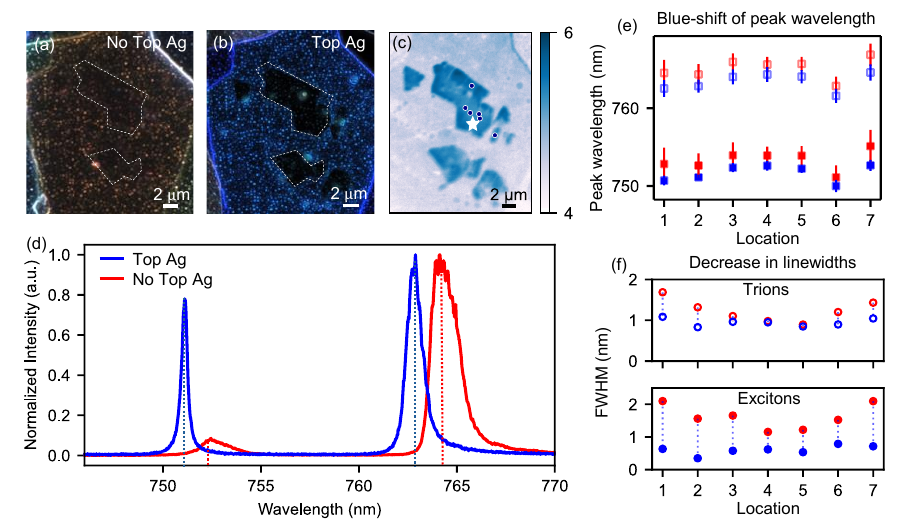}}
\caption{Change in emission spectra of excitons and trions post Ag growth (a,b) Dark-field microscope image of the stack (a) before Ag growth and (b) after Ag growth. (c) PL map obtained with green laser excitation of the flake with points corresponding to the measurements of change in peak wavelength and linewidths with the star indicating the point where the representative spectra (d) is measured. (d) Emission profile of excitons and trions before (red) and after (blue) Ag growth (e) Consistent blue-shift in the peak wavelength of trions (open markers) and excitons (filled markers) post Ag deposition at 7 different points indicating a cavity-induced wavelength tuning effect. (f) Distribution of reduction in linewidths of excitons and trions obtained across the flake at 7 different points. }\label{Figure2}
\end{figure}

We sequentially perform measurements to measure the impact of cavity integration on radiative processes. The first round of data were taken without the top layer of silver and the second round of data were taken after the deposition of the top layer of silver. Dark-field microscopy images of the sample are shown in Figures \ref{Figure2}a,b before and after the top layer of Ag was deposited. The MoSe$_2$ monolayers are outlined. A spatially resolved photoluminescence (PL) map of the sample using a 520 nm excitation laser diode is shown in Figure \ref{Figure2}c with a representative emission spectrum at the point labeled by the star in Figure \ref{Figure2}d. We find that the exciton and trion peaks show linewidth narrowing as well as a blue-shift of the peak wavelength with the same excitation power. This is consistent across several different points across the sample (denoted by black dots in Figure \ref{Figure2}c and labeled in SI Figure 3). We attribute the linewidth narrowing to an increase in the exciton and trion lifetimes. We fit the PL spectra to a Lorentzian function to extract the exciton and trion energies and linewidths. The sample shows an average exciton linewidth decrease from $3.53 \pm0.82$ meV to $1.32\pm0.31$ meV and an average trion linewidth decrease of $2.61\pm0.58$ meV to $2.01\pm0.20$ meV. The full table of spectra fits from each location are included in the SI. The trion linewidth narrowing is less dramatic than the exciton linewidth likely due to more non-radiative decay pathways \cite{gillespie_optical_2024}. We also observe a consistent blue-shift of $3.8 \pm 0.8$ meV and $3.6\pm0.9$ meV for the exciton and trion peak wavelengths after the top layer of Ag is deposited. This simultaneous decrease in exciton line width and blue shift of the peak energy suggests a cavity-induced cooperative Lamb shift of excitons\cite{horng_engineering_2019} \cite{fang_control_2019}. Previous studies of exciton linewidth and energy modulation by a single DBR mirror \cite{horng_engineering_2019} or dielectric environment \cite{ren_control_2023} demonstrate PL linewidth narrowing from $\sim4.5$meV to $\sim2$meV and Lamb shifts of $\sim1$meV. The plasmonic cavity yields a significantly larger Lamb shift in our case compared to previously reported values due to the tightly confined and suppressed emission of the excitons. 

\begin{figure}[H]
\centerline{\includegraphics{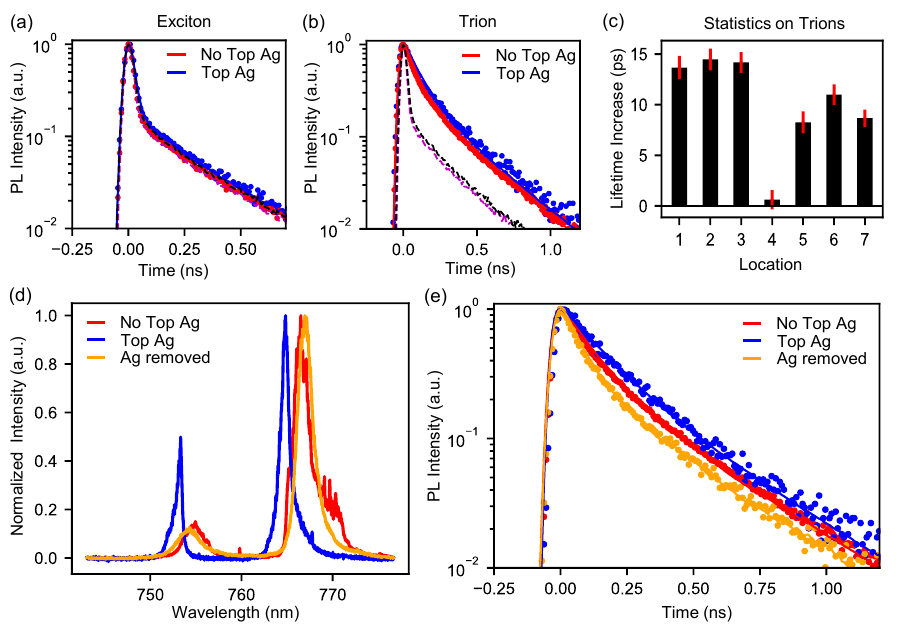}}
\caption{Effect of Ag growth and removal (a) Increase in exciton lifetime overlaid with the IRF (dotted) indicating the pre-Ag lifetime of excitons is below the detection limit of the fast APD. (b)Increase in trion lifetime (c) Distribution of increase in lifetime of trions obtained across the flake at 7 different points. (d) Peak wavelength and linewidth recovered after removing top Ag layer. (e) Change in trion lifetime recorded pre-Ag growth, post-Ag growth and Ag-etch.}\label{Figure3}
\end{figure}

To confirm that the addition of the top layer of Ag does indeed increase the lifetime of excitons and trions, we performed lifetime measurements of each species using time-correlated single photon counting (see Methods section for more details). The results of the lifetime measurements are presented in Figure \ref{Figure3}. Figures \ref{Figure3}a and b show the exciton and trion lifetimes, respectively, before and after the addition of the top Ag. The lifetime data are convolved with the instrument response function (IRF), which is overlaid in each plot. We note that before the top layer of Ag was deposited, the exciton lifetime coincides with the IRF, indicating that the lifetimes are below the detection limit of the instruments. However, after the top layer of Ag was deposited, the exciton lifetimes rose above the IRF, with an average lifetime of $\sim 9$ps. The trion lifetime pre-Ag was, on average, 59 ps, and after the top Ag was deposited, we measured an average lifetime increase of $\sim 10$ ps. The lifetime increase for each point is shown in Figure \ref{Figure3}c. 

Moreover, we demonstrate that the linewidth narrowing and lifetime increase can be modulated. We etch away the top Ag and perform the measurements again. Figure \ref{Figure3}d,e show the emission spectra and trion lifetime at all three stages of the experiment. We observe that when the Ag is removed, the emission spectrum returns to that of the spectrum taken before the Ag, with similar linewidths and peak energy. The trion lifetime also decreases after the top Ag is removed. We note that the degree of lifetime enhancement is less than predicted by theory and simulation. We speculate that this is limited by disorder in the sample and other non-radiative recombination pathways. These measurements confirm that our observed modulation of exciton linewidth and lifetime arise purely due to interaction with the cavity structure.

In summary, we fabricated a tightly-confined planar cavity using exfoliated mica as a dielectric, lattice-matched layer for silver and demonstrated that placing monolayer MoSe$_2$ in a planar silver cavity results in a significant decrease in exciton and trion linewidths, corresponding to an increase in lifetime. This is confirmed by lifetime measurements, which show that the exciton lifetime increases above the instrument IRF with the addition of the top silver, and an increase in the trion lifetime. We confirm that this effect is due to the addition of the top silver by etching away the silver and re-performing the measurements, which shows that the spectra returns to its original shape and the lifetime decreases again. Additionally, we find a consistent blue-shift of the exciton and trion peak wavelengths, which suggests a cavity-induced cooperative Lamb shift. 

Future studies include studying the gate-dependent behavior of the exciton and trion/polaron emission within the tightly confined cavity. Prolonging the valley-polarization lifetime of trions can significantly improve their suitability for valleytronic devices, enabling applications such as electrically controlled logic gates in quantum computing\cite{pawlowski_valley_2021}. The ability to control and extend the lifetimes of excitons and trions can lead to enhanced optical gain and lasing capabilities\cite{duan_valley-addressable_2023}. This versatility in manipulating excitonic emissions offers a promising platform for developing novel optoelectronic devices with tunable properties such as frequency, intensity, and polarization. While the planar Fabry-Pérot cavity suppresses the emission of in-plane dipoles, it enhances the emission of out-of-plane dipoles \cite{chen_engineering_2022}. The dark exciton in MoSe$_2$ is predicted to be a higher energy excitation in contrast to WSe$_2$ where it is the lowest energy excitation\cite{robert_measurement_2020}. The dark exciton has an out-of-plane dipole moment which can couple to surface plasmon polaritons (SPP) in the nearby silver \cite{zhou_probing_2017}. Thus, integrating the Fabry-Pérot cavity which suppresses the bright exciton with nearby structures that can scatter dark excitons coupled to SPPs for far-field detection can allow for the detection of dark excitons in MoSe$_2$. Finally, MoSe$_2$/WSe$_2$ hetero-bilayers host long-lived interlayer excitons which also have an in-plane optical dipole moment and are predicted to form a Bose-Einstein condensate (BEC)\cite{wang_evidence_2019}\cite{ma_strongly_2021} \cite{qi_thermodynamic_2023}\cite{qi_perfect_2025}\cite{nguyen_perfect_2025}. Integrating these systems with the tightly confined metallic Fabry-Pérot cavities can further prolong the exciton lifetime and allow further studies of BEC physics.

\section{Methods} \label{Methods}
\subsection{Sample Preparation}
\subsubsection{Single-crystalline silver substrate}
We use high-grade muscovite mica as the lattice-matched growth substrate for single-crystalline silver with sub-nanometer surface roughness\cite{park_singlecrystalline_2012}. The mica substrate is first cleaved directly before loading into the thermal evaporation chamber (Angstrom Nexdep). This ensures a fresh and clean surface for silver depositon. The substrate is heated to 350 C to overcome the large energy difference between the Ag and mica and prevent the formation of 3D islands \cite{park_singlecrystalline_2012}. We deposit Ag at a rate of 1.8 nm/s to ensure a smooth and continuous film. 
\subsubsection{TMD heterostructure}
Monolayer MoSe$_{2}$ (HQ Graphene), hBN, and mica are mechanically exfoliated onto SiO$_{2}$/Si substrates. Flakes of desirable thickness are identified optically and verified via atomic force microscopy. The heterostructure is assembled with a dry-transfer process using a PC/PDMS stamp and released onto the single-crystalline silver substrate. After stacking, the PC is dissolved in chloroform.
\subsubsection{Silver cavity fabrication}
Following the first round of measurements, we deposit 40nm of silver at 0.5 nm/s using thermal evaporation on top of the sample without any substrate heating. This silver film is expected to be poly-crystalline.
\subsubsection{Removal of top silver}
The back-etch of Ag was performed with Ar gas (15 sccm flow) at 2 mTorr pressure at 300 W ICP bias for 2.5 seconds. 
\subsection{Simulations}
\noindent FDTD simulations were done using Ansys Lumerical by placing an in-plane dipole at the center of the monolayer. The full stack design (also shown in SI Figure S1) including the Ag cavity is designed with a mica substrate, a layer of single-crystal Ag (500 nm), bottom hBN (48 nm), MoSe$_2$ flake, top hBN (variable), mica (variable), top layer of poly-crystalline Ag (40 nm). A parameter sweep is performed to calculate the ideal thicknesses of top hBN and mica based on the Purcell factor calculation with and without the Ag cavity (top and bottom Ag layers). The maximum Purcell suppression is used as the figure of merit to get the final thicknesses of hBN and mica.
\subsection{Optical Measurements}
\noindent The sample was addressed in a home-built confocal 4f microscope. Photoluminescence spectra was obtained using a 520nm diode (10$\mu$W of power and sub-micron spotsize) and a Horiba 550 spectrometer (Synapse Plus camera and a 1200g/mm grating). Lifetime measurements were obtained using the time-correlated single photon counting technique\cite{buschmann_characterization_2013}. Our single photon detector is a fast APD from MPD (PD-050-CTC-FC), our time tagger is the QuTag from QuTools. A Ti$\colon$\!Saph oscillator tuned to 717 nm, with a broad cleanup filter (Semrock 720BP), was used to excite the sample with an average power of 5 $\mu$W and repetition rate of 76 MHz. Additionally, a Semrock bandpass filter (775/46) in combination with tilting a 750/5 (exciton) or 770/5 (trion) was used to spectrally separate the two species. The IRF was obtained by measuring the reflected laser light with an OD 8.5 filter and a 758 shortpass filter to extinguish sample photoluminescence. All measurements were done at 4K in a Montana S100 Cryostation with a 0.9 NA Cryo-Objective. 

\begin{acknowledgement}
G. H. Chen acknowledges funding from the National Science Foundation Graduate Research Fellowship under Grant No. DGE 2140743. A. Addhya acknowledges funding from Q-NEXT, supported by the U.S. Department of Energy, Office of Science, National Quantum Information Science Research Centers and Kadanoff-Rice fellowship (NSF DMR-2011854). I.N. Hammock acknowledges funding the National Science Foundation's Quantum Leap Challenge Institute
for Hybrid Quantum Architectures and Networks (HQAN) (NSF OMA-
2016136). This work made use of the Pritzker Nanofabrication Facility (Soft and Hybrid Nanotechnology Experimental Resource, NSF ECCS-2025633) at the University of Chicago. This work was partially supported by the University of Chicago Materials Research Science and Engineering Center, which is funded by the National Science Foundation under award number DMR-2011854.

\end{acknowledgement}

\begin{suppinfo}
Additional device details, characterization, and measurement data are included in the Supporting Information.

\end{suppinfo}

\bibliography{referencesz}

\end{document}